\documentclass[11pt]{article}
\usepackage{moriond,epsfig}

\usepackage{mathrsfs}
\usepackage{latexsym}
\usepackage{amsfonts}
\usepackage{amssymb}

\def\be{\begin{equation}}
\def\ee{\end{equation}}
\def\bea{\begin{eqnarray}}
\def\eea{\end{eqnarray}}
\def\Gammabol{{\stackrel{\circ}{\Gamma}}{}}

\def\Rbol{{\stackrel{\circ}{R}}{}}

\def\Gammabol{{\stackrel{\circ}{\Gamma}}{}}
\def\Tbol{{\stackrel{\circ}{T}}{}}

\def\Gammaw{{\stackrel{\bullet}{\Gamma}}{}}

\def\Rw{{\stackrel{\bullet}{R}}{}}

\def\Tw{{\stackrel{\bullet}{T}}{}}
\def\Kw{{\stackrel{\bullet}{K}}{}}

\def\onehalf{{\textstyle{\frac{1}{2}}}}

\begin{document}
\vspace*{4cm}
\title{IN SEARCH OF THE SPACETIME TORSION}

\author{J. G. PEREIRA}

\address{Instituto de F\'{\i}sica Te\'orica, Universidade Estadual Paulista \\
Rua Pamplona 145, 01405-900 S\~ao Paulo, Brazil}

\maketitle
\abstracts{Whether torsion plays or not a role in the description of the 
gravitational interaction is a problem that can only be solved by experiment. This 
is, however, a difficult task: since there are different possible interpretations 
for torsion, there is no a model--independent way to look for it. In these notes, 
two different possibilities will be reviewed, their consistency analyzed, and the 
corresponding experimental outputs briefly discussed.}

%%%%%%%%%%%%%%%%%%%%%%
\section{Gravitation and Universality}

Gravitation has a quite peculiar property: particles with different masses and 
different compositions feel it in such a way that all of them acquire the same 
acceleration and, given the same initial conditions, follow the same path. Such 
universality of response --- usually referred to as {\it universality of free  
fall} --- is the most fundamental characteristic of the gravitational 
interaction.\cite{mtw} It is unique, peculiar to gravitation: no other basic 
interaction of nature has it. Universality of free fall is usually identified with 
the weak equivalence principle, which establishes the equality between inertial 
and gravitational masses. In fact, in order to move along the same trajectory, the 
motion of different particles must be independent of their masses, which have then 
to be canceled out from the equations of motion. Since this cancellation can only 
be made when the inertial and gravitational masses coincide, this coincidence 
naturally implies universality.

Einstein's general relativity is a theory fundamentally based on the weak 
equivalence principle. In fact, to comply with universality, the presence of a 
gravitational field is supposed to produce {\em curvature} in spacetime, the 
gravitational interaction being achieved by letting (spinless) particles to follow 
the {\it geodesics} of the curved spacetime. In general relativity, therefore, 
geometry replaces the concept of gravitational force, and the trajectories are 
determined, not by force equations, but by geodesics. It is important to emphasize 
that only a universal interaction can be described by such a geometrization, in 
which the responsibility for describing the interaction is transferred to 
spacetime. It is also important to remark that, in the eventual lack of 
universality, the geometrical description of general relativity would break 
down.\cite{wep}

The fundamental connection of general relativity is the Christoffel connection, a 
Lorentz-valued connection written in a coordinate basis.\cite{livro} In terms of 
the spacetime\hskip 0.08cm\footnote{We use the Greek alphabet $\mu, \nu, \rho, 
\dots = 0, 1, 2, 3$ to denote spacetime indices.} metric $g_{\mu \nu}$, it is 
written as
\be
\Gammabol^{\rho}{}_{\mu\nu} = \onehalf \, g^{\rho \lambda} (\partial_\mu 
g_{\lambda \nu} +
\partial_\nu g_{\lambda \mu} - \partial_\lambda g_{\mu \nu}).
\ee
It is a connection with vanishing torsion, $\Tbol^{\rho}{}_{\nu\mu} = 0$, but non-
vanishing curvature, $\Rbol^\rho{}_{\lambda\nu\mu} \neq 0$. In terms of this 
connection, the equation of motion of a test particle is given by the geodesic 
equation
\be
\frac{d u_\mu}{d s} -
{\stackrel{\circ}{\Gamma}}{}^\theta{}_{\mu \nu} \; u_\theta \; u^\nu = 0,
\ee
which says that the particle four-acceleration vanishes identically. This property 
reveals the absence of the concept of gravitational force, a basic characteristic 
of the geometric description.

%%%%%%%%%%%%%%%%%%%%%%%%%%%%%%
\section{What About Torsion?}

A general Lorentz connection has two fundamental properties: curvature and 
torsion.\cite{koba} Why should then matter produce {\it only curvature}? Was 
Einstein wrong when he made this assumption by choosing the Christoffel 
connection? Does torsion play any role in gravitation? Cartan was the first to ask 
these questions, soon after the advent of general relativity. As a possible 
answer, a new theory was formulated, called {Einstein--Cartan theory},\cite{ect} 
in which the Christoffel connection was replaced by a general connection 
presenting both curvature and torsion. 

The main idea behind the Einstein--Cartan construction is the fact that, at the 
microscopic level, matter is represented by elementary particles, which in turn 
are characterized by mass (that is, energy and momentum) and spin. If one adopts 
the same {\it geometrical spirit of general relativity}, not only mass but also 
spin should be source of gravitation at this level. Like in general relativity, 
energy and momentum should appear as source of curvature, whereas spin should 
appear as source of torsion. This means essentially that, in this theory, 
curvature and torsion represent independent degrees of freedom of gravity. As a 
consequence, there might exist new physics associated to torsion. Of course, at 
the macroscopic level, where spins vanish, the Einstein--Cartan theory coincides 
with general relativity. At the microscopic level, however, where spins are 
relevant, it shows different results from general relativity. If this is 
interpretation is correct, Einstein made a mistake when he did not include torsion 
in general relativity.

Because of the Einstein--Cartan theory, there is a widespread belief that torsion 
is intimately related to spin, and consequently important only at the microscopic 
level. This belief, however, is not fully justified: in addition to lack 
experimental support, it is based on a very particular model for gravitation, 
which is well known to present several consistency problems. For example, it is 
not consistent with the strong equivalence principle.\cite{ap0} Another relevant 
problem is that, when used to describe the interaction of the electromagnetic 
field with gravitation, the Einstein--Cartan coupling prescription violates the 
U(1) gauge invariance of Maxwell's theory.\cite{hehl} This last problem is usually 
circumvented by {\it postulating} that the electromagnetic field does not couple 
to torsion.\cite{postulate} This ``solution'', however, is quite 
unreasonable.\cite{vector} The purpose of these notes is to call the attention to 
another possible interpretation for torsion,\cite{ap0} which is consistent and 
does not present the above mentioned problems. This solution is based on a 
different model for gravitation, known as the {\it teleparallel gravity equivalent 
of general relativity},\cite{ap1} or simply {\it teleparallel gravity}. 

%%%%%%%%%%%%%%%%%%%%%%%%%%%%%%%%%%%%%%%%%%%%%%%
\section{A Glimpse on Teleparallel Gravity}

Teleparallel gravity corresponds to a gauge theory for the translation group, with 
the field strength given by the torsion tensor. Its main difference in relation to 
general relativity is the connection field: instead of Christoffel, the 
fundamental connection of teleparallel gravity is the so called Weit\-zen\-b\"ock 
connection. In terms of the tetrad~\footnote{We use the Latin alphabet $a, b, c, 
\dots = 0, 1, 2, 3$ to denote algebraic indices related to the tangent Minkowski 
spaces. These indices are raised and lowered with the Minkowski metric 
$\eta_{ab}$, whereas the spacetime indices are raised and lowered with the metric 
$g_{\mu \nu} = \eta_{ab} \, h^a{}_\mu \, h^b{}_\nu$.} $h^a{}_\mu$, it is written 
as
\be
\Gammaw^{\rho}{}_{\mu\nu} = h_a{}^\rho \, \partial_\nu h^a{}_\mu.
\ee
In contrast to Christoffel, it is a connection with non-vanishing torsion, 
$\Tw^{\rho}{}_{\nu\mu} \neq 0$, but vanishing curvature, $\Rw^{\lambda 
\rho}{}_{\nu\mu} = 0$. The Weitzenb\"ock and the Christoffel connections are 
related by
\be
\Gammaw^{\rho}{}_{\mu\nu} = \Gammabol^{\rho}{}_{\mu\nu} + \Kw^{\rho}{}_{\mu\nu},
\label{rela}
\ee
where
\be
\Kw^{\rho}{}_{\mu\nu} = \onehalf \, (\Tw_\mu{}^\rho{}_\nu + \Tw_\nu{}^\rho{}_\mu - 
\Tw^{\rho}{}_{\mu\nu})
\ee
is the contortion tensor.

In teleparallel gravity, the equation of motion of a test particle is given by the 
{\it force equation}~\cite{paper1}
\be
\frac{d u_\mu}{d s} -
\Gammaw^\theta{}_{\mu \nu} \; u_\theta \; u^\nu =
\Tw^\theta{}_{\mu \nu} \; u_\theta \, u^\nu,
\label{forceq}
\ee
with torsion playing the role of {\it gravitational force}. It is similar to the 
Lorentz force e\-qua\-tion of electrodynamics, a property related to the fact 
that, like Maxwell's theory, teleparallel gravity is also a gauge theory. Using 
expression (\ref{rela}), the force equation (\ref{forceq}) can be rewritten in 
terms of the Christoffel connection, in which case it reduces to the geodesic 
equation of general relativity:
\be
\frac{d u_\mu}{d s} -
{\stackrel{\circ}{\Gamma}}{}^\theta{}_{\mu \nu} \; u_\theta \; u^\nu = 0.
\label{geobis}
\ee
The force equation (\ref{forceq}) of teleparallel gravity and the geodesic 
equation (\ref{geobis}) of general relativity, therefore, describe the same 
physical trajectory. This means that the gravitational interaction has two {\em 
equivalent} descriptions: one in terms of curvature, and another in terms of 
torsion.\cite{equiva} Although equivalent, however, there are conceptual 
differences between these two descriptions. In general relativity, curvature is 
used to {\it geometrize} the gravitational interaction. In teleparallel gravity, 
on the other hand, torsion accounts for gravitation, not by geometrizing the 
interaction, but by acting as a {\it force}. As a consequence, there are no 
geodesics in teleparallel gravity, but only force equations, quite analogous to 
the Lorentz force e\-qua\-tion of electrodynamics.

One may wonder why gravitation has two equivalent descriptions. This duplicity is 
related precisely to that peculiarity: universality. Like the other fundamental 
interactions of nature, gravitation can be described in terms of a gauge theory --
- just teleparallel gravity. Universality of free fall, on the other hand, makes 
it possible a second, geometrized description, based on the weak equivalence 
principle --- just general relativity. As the sole universal interaction, it is 
the only one to allow also a geometrical interpretation, and hence two alternative 
descriptions. From this point of view, curvature and torsion are simply 
alternative ways of describing the gravitational field,\cite{aap} and consequently 
related to the same degrees of freedom of gravity. If this interpretation is 
correct, Einstein was right when he did not include torsion in general relativity.

%%%%%%%%%%%%%%%%%%%%%%%%%
\section{Conclusions}

According to the Einstein--Cartan theory, as well as to other generalizations of 
general relativity, torsion represents additional degrees of freedom of gravity. 
As a consequence, new physical phenomena associated to its presence are predicted 
to exist. With this point of view in mind, there has been recently a proposal to 
look for these new phenomena using the Gravity Probe B data.\cite{mao} The idea is 
that, if torsion is able to couple to spin, consistency arguments require that it 
might also be able to couple to rotation --- that is, to orbital angular momentum. 
The data of Gravity Probe B could then be used to look for signs of this coupling. 
On the other hand, from the point of view of teleparallel gravity, torsion does 
not represent additional degrees of freedom, but simply an alternative to 
curvature in the description of gravitation. In this case, there are no new 
physical effects associated with torsion. According to teleparallel gravity, 
therefore, torsion has already been detected: it is the responsible for all known 
gravitational effects, including the physics of the solar system, which can be 
reinterpreted in terms of a force equation, with torsion playing the role of 
gravitational force.

Which of these interpretations is the correct one? From the theoretical point of 
view, we can say that the teleparallel interpretation presents several conceptual 
advantages in relation to the Einstein--Cartan theory: it is consistent with the 
strong equivalence prin\-ciple,\cite{ap0} and when applied to describe the 
interaction of the electromagnetic field with gravitation, it does not violate the 
U(1) gauge invariance of Maxwell's theory.\cite{vector} From the experimental 
point of view, on the other hand, at least up to now, there are no evidences for 
new physics associated with torsion. We could then say that the existing 
experimental data favor the teleparallel point of view, and consequently general 
relativity. However, due to the weakness of the gravitational interaction, no 
experimental data exist on the coupling of the spin of the fundamental particles 
to gravitation. For this reason, in spite of the conceptual soundness of the 
teleparallel interpretation, we prefer to say that a definitive answer can only be 
achieved by further experiments.

%%%%%%%%%%%%%%%%%%%%%%%%%%
\section*{Acknowledgments}
The author would like to thank R.\ Aldrovandi and H.\ I.\ Arcos for useful 
discussions. He would like to thank also FAPESP, CNPq and CAPES for partial 
financial support.

%%%%%%%%%%%%%%%%%%%%%%%%%%%

\end{document}